\documentclass[final]{aipproc}
\usepackage{amssymb}
\usepackage{amsmath}
\usepackage{ifpdf}

\layoutstyle{6x9}

\newcommand\beq{\begin{equation}}
\newcommand\eeq{\end{equation}}

\title{When Stars Collide}


\author{E.~Glebbeek}{
  address={Sterrekundig Instituut Utrecht, P.O. Box 80000, 3508 TA Utrecht, The Netherlands}
}

\author{O.~R.~Pols}{
  address={Sterrekundig Instituut Utrecht, P.O. Box 80000, 3508 TA Utrecht, The Netherlands}
}

\begin{document}

\begin{abstract}
When two stars collide and merge they form a new star that can stand out against
the background population in a starcluster as a blue straggler. In so called
collision runaways many stars can merge and may form a very massive star that
eventually forms an intermediate mass blackhole. We have performed detailed 
evolution calculations of merger remnants from collisions between main 
sequence stars, both for lower mass stars and higher mass stars. 
These stars can be significantly brighter than ordinary stars of the same
mass due to their increased helium abundance. Simplified treatments ignoring
this effect give incorrect predictions for the collision product lifetime and 
evolution in the Hertzsprung-Russell diagram.
\end{abstract}

\maketitle

\section{Introduction}
In star clusters stars can experience close encounters with other stars,
which in some cases can lead to the collision and merger of two or more
stars. This is a possible formation mechanism for blue straggler stars.

\begin{figure}
\ifpdf
   \includegraphics[width=\textwidth]{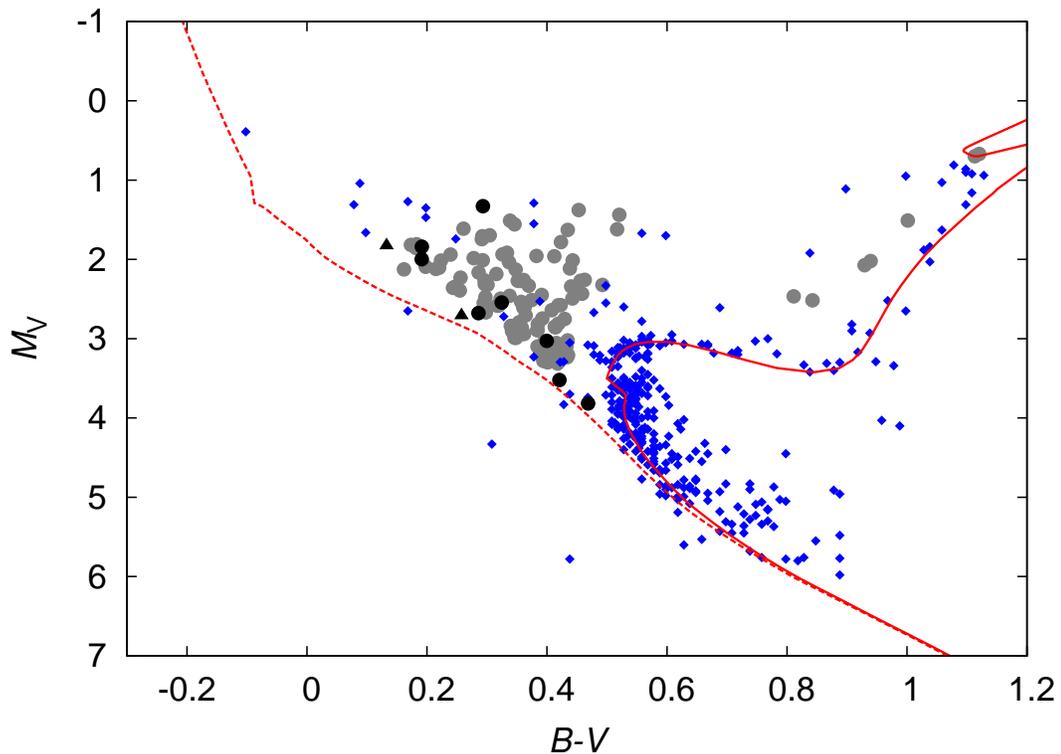}
\else
   \includegraphics[angle=270,width=\textwidth]{m67_hrd-4}
\fi
\caption{Colour-magnitude diagram of the old open cluster M67 ($\diamond$)
from \citep{article:montgomery_m67} with the locations of our merger 
remnants at $4 \mathrm{Gyr}$ overplotted ($\bullet$, $\blacktriangle$). Collision
products from the M67 simulation are plotted in black, collision products from
our collision grid are plotted in grey. The triangles indicate the position of
the remnants from a double collision.
Plotted isochrones are for the zero-age main sequence (dashed) and
$4 \mathrm{Gyr}$ (solid). For discussion of the isochrones and calculation of 
the $B-V$ colours, see \citet{article:pols_evmodels}}
\label{fig:hrd_m67}
\end{figure}

Blue stragglers are stars that appear on the extension of 
the main sequence in the colour-magnitude diagram (CMD) of star clusters 
(\citet{article:piotto_freq_bss} and figure \ref{fig:hrd_m67}).
In the past, various mechanisms for their formation have been proposed 
(see \emph{e.~g.~} \citet{article:livio_blue_stragglers} for a list),
but the most commonly accepted explanation is that they are stars that have
gained mass long after they were formed, either through mass transfer in a
close binary or by merging two stars. Such a merger can in turn be the result
of normal binary evolution or of a collision with another star. These 
mechanisms are thought to operate and produce blue stragglers in different 
regions in a cluster \citep{article:davies_bss_formatio}. We have studied mergers that result from
collisions.

In some cases a sequence of collisions in the centre of a cluster can lead to 
a runaway where many stars merge together 
\citep{article:portegies_zwart_intermediate_mass_blackhole}. Such runaways
may lead to the formation of very massive stars and ultimately intermediate
mass blackholes. We have studied the evolution of the outcome of a single
collision.

One of the goals of the MODEST collaboration \citep{article:modest1}
is to study the evolution of stellar mergers by combining stellar evolution,
stellar hydrodynamics and stellar dynamics in one software framework. This
work is the first result from combining stellar hydrodynamics and stellar
evolution in one framework.

\section{Structure of collision products}
During a collision the potential energy of the two parent stars is released 
and converted to thermal energy. This means that the collision product
initially has a lot of excess thermal energy that needs to be lost by
gravitational contraction. In this sense, the collision product resembles a
pre-main sequence star.

Unlike pre-main sequence stars, however, collision products are not
convective and are not well mixed \citep{article:sills_on_axis}. In many 
simplified descriptions of stellar mergers, the merger remnants are either 
treated as
normal zero-age main sequence stars with the same total mass, or the two stars 
are assumed to mix homogeneously during the collision 
\citep{article:tout_evolution_models}. Neither of
these is a very good approximation and the resulting evolution can be very
different from a more detailed approach.

In a realistic collision, the collision product is expected to have gained a
lot of angular momentum and should spin rapidly. Some of this excess angular
momentum needs to be lost before the star can contract to a main sequence
position. The mechanism for this is not well understood although magnetic
fields are thought to play a role \citep{article:sills_angular_momentum}. Rapid rotation can 
lead to extra mixing and affect the evolution of the stars 
\citep{article:heger_rotation_1, article:pinsonneault_rotation}.
Because the mechanism by which angular momentum is lost is not well understood 
we have not considered the effect of rotation on the results presented here,
but we will address this in future work.

We have performed detailed evolution calculations for remnants of collisions
involving low mass stars as well as high mass stars. The evolution tracks were
calculated with the stellar evolution code originally developed by Eggleton
\citep{article:eggleton_evlowmass, article:pols_approxphys}, the structure of 
collision products were calculated in collaboration with E. Gaburov and 
S. Portegies Zwart using
smooth particle hydrodynamics for the more massive stars, and using the
parametric code of Lombardi \citep{article:lombardi_mmas} for the lower mass 
stars.

\section{Collisions between low mass stars: blue stragglers}
\begin{figure}
\ifpdf
   \includegraphics[width=\textwidth]{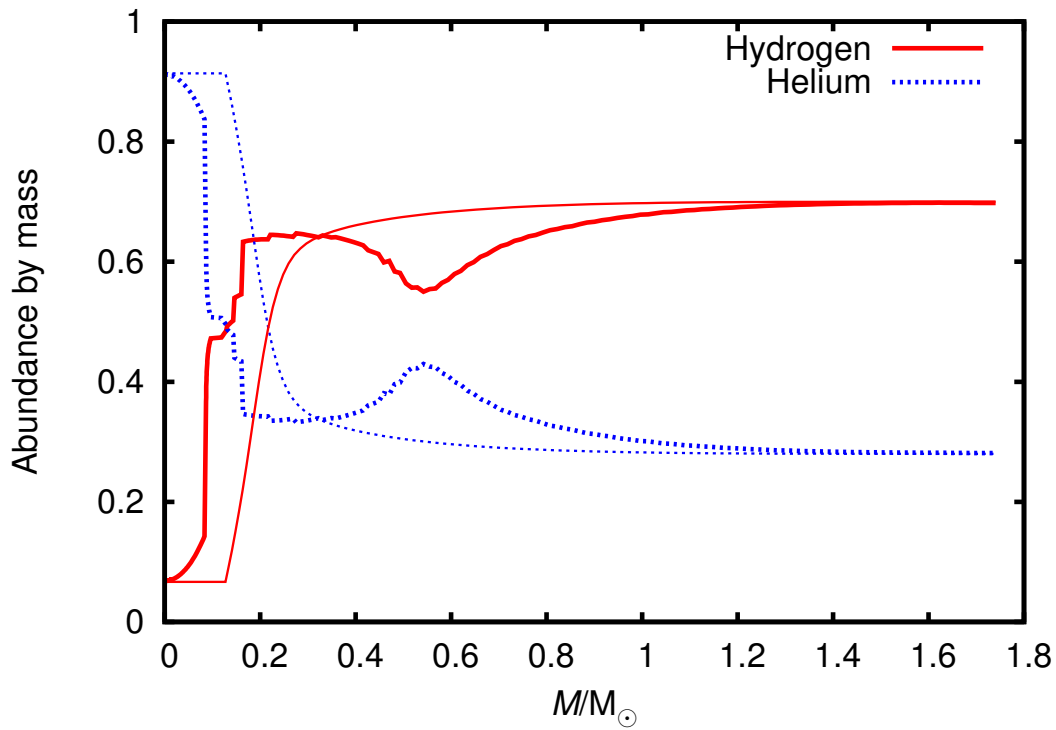}
\else
   \includegraphics[angle=270,width=\textwidth]{2321_comp_merger-1}
\fi
\caption{Composition profile of the merger remnant from a collision between
a $1.29$ and a $0.59 \mathrm{M}_\odot$ star (thick line), compared to that of a normal
star with the same core hydrogen content (thin line). Shown profiles are for
hydrogen (solid line) and helium (dashed line).}
\label{fig:merger_composition}
\end{figure}
As noted, collisions between low mass stars ($M_1+M_2 \lesssim 2 \mathrm{M}_\odot$) are
interesting in the context of blue straggler formation in globular clusters and
old open clusters. 

We have made detailed models of collision products for collisions found in 
an $N$-body simulation of M67 \citep{article:hurley_m67} and evolved these to $4 \mathrm{Gyr}$,
the age of M67. More details for this will be given in a forthcoming paper
\citep{article:glebbeek_pols_in_prep1}.

\begin{figure}
\ifpdf
   \includegraphics[width=\textwidth]{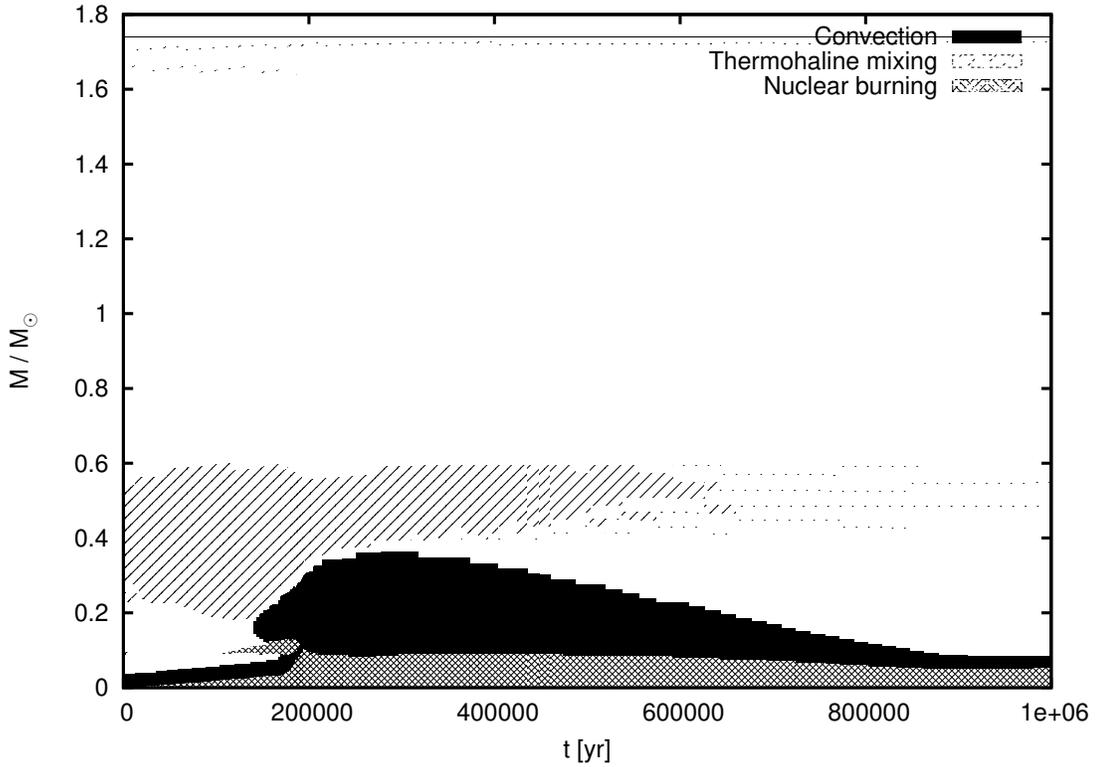}
\else
   \includegraphics[angle=270,width=\textwidth]{2321_kippenhahn}
\fi
\caption{Kippenhahn plot of the early evolution of the collision product from
figure \ref{fig:merger_composition}. Black regions are convective, while
hatched regions are unstable to thermohaline mixing. Nuclear burning regions are
indicated by the cross hatching.}
\label{fig:2321_kippenhahn}
\end{figure}
At $t = 3960\mathrm{Myr}$ a collision occured in the simulation between a $1.29 \mathrm{M}_\odot$
primary and a $0.59 \mathrm{M}_\odot$ secondary star. The primary stars was close to
the end of its main sequence. After the collision the collision product had 
the composition profile
shown in figure \ref{fig:merger_composition}. A large portion of the core from
the primary sank to the centre of the collision product, leading to a 
hydrogen-poor region below $M=0.1 \mathrm{M}_\odot$. Above this region and below
$0.4 \mathrm{M}_\odot$ most of the material comes from the core of the secondary star.
Material from the core of the primary lead to an increase in the helium
abundance between $M=0.4 \mathrm{M}_\odot$  and $M=0.6 \mathrm{M}_\odot$. Material further out
has the normal unprocessed composition.
The interior is mixed during
the early evolution of the collision product by three processes. The sharp
increase in the hydrogen abundance at $M=0.1 \mathrm{M}_\odot$ leads to a peak in the 
nuclear reaction rate at this location, which leads to the development of a 
hydrogen burning shell (see figure \ref{fig:2321_kippenhahn}). The hydrogen 
burning shell drives a convection zone that eventually connects to the
convective core, mixing the inner $0.4 \mathrm{M}_\odot$. The region above this is
unstable to thermohaline mixing \citep{article:kippenhahn_thermohalinemixing,
article:stancliffe_thmixing} due to the 
molecular weight inversion (see figure \ref{fig:2321_kippenhahn}). 

\begin{figure}
\ifpdf
   \includegraphics[width=\textwidth]{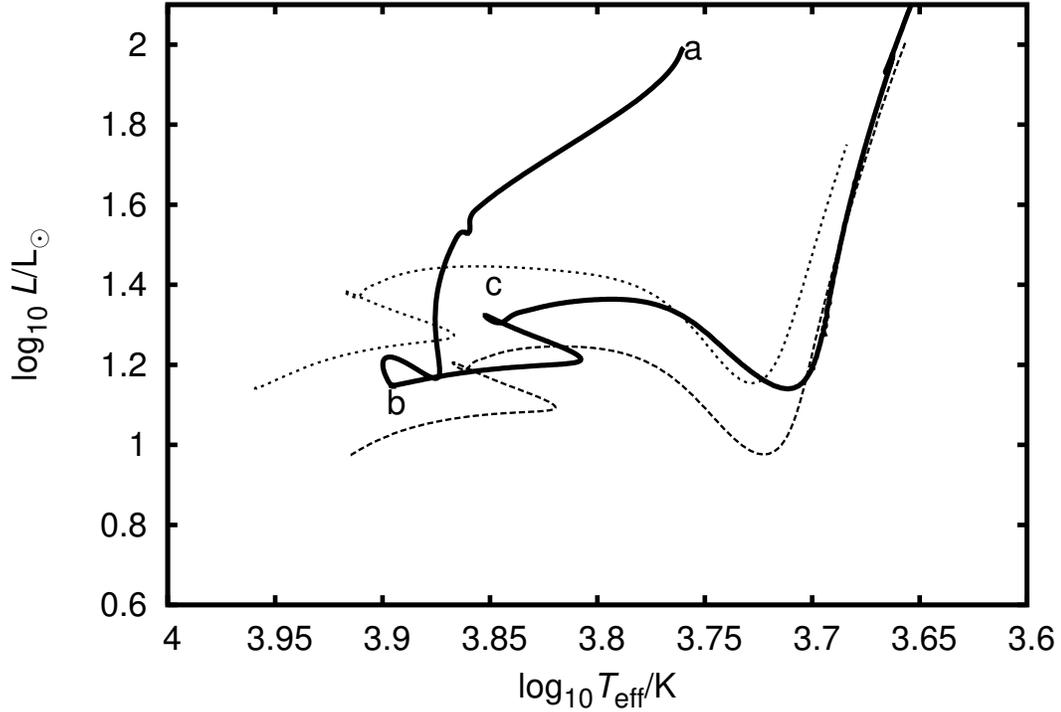}
\else
   \includegraphics[angle=270,width=\textwidth]{hrd_2321-4}
\fi
\caption{Evolution track of a merger remnant from figure
\ref{fig:merger_composition} compared to that of a normal star born with the
same mass (dashed) and a homogeneously mixed version of the collision product
(dotted). The collision product starts contracting at point (a) and reaches
the main sequence at point (b) after $2 \mathrm{Myr}$ and the end of the
main seqeuence at (c) after $0.6 \mathrm{Gyr}$.}
\label{fig:hrd_lowmass}
\end{figure}

By the time the star has reached thermal equilibrium the inner $0.6 \mathrm{M}_\odot$ 
has been mixed and is helium enhanced compared to the material of the 
envelope. As a consequence, the mean molecular weight in the star is
increased compared to that of a normal star. This affects the luminosity
according to the scaling relation \citep{book:kippenhahn_weigert}
\beq
L_\mathrm{merger} = L_\mathrm{ms} \frac{\mu_\mathrm{merger}^4}{\mu_\mathrm{ms}^4}.
\eeq
The star's temperature is not strongly affected because the volume averaged
opacity is not significantly changed. This is because the hydrogen envelope
comprises most of the star's volume. As a result, the collision product
is brighter than a normal star of its mass but not as blue as it would
have been if the star had been homogeneously mixed (figure \ref{fig:hrd_lowmass}).

The hydrogen abundance in the core has increased by the mixing, extending the
lifetime: the star is rejuvenated.

In addition to the mergers found in the $N$-body simulation we have computed
a number of models to investigate the parameter space $M_1$, $M_2$ and 
$t_\mathrm{collision}$ for the merger remnants \citep{article:glebbeek_pols_in_prep2}. The results 
of these calculations are shown as cicles in figure \ref{fig:hrd_m67}. The
models from our grid nicely fill the blue straggler region in the observed
CMD, confirming that merger remnants do look like blue stragglers. The absense
of models around the blue and red boundaries of the blue straggler region are
an artefact due to the limited range in $M_1+M_2$ of our grid.

\section{Collisions between high mass stars}
\begin{figure}
\ifpdf
   \includegraphics[width=\textwidth]{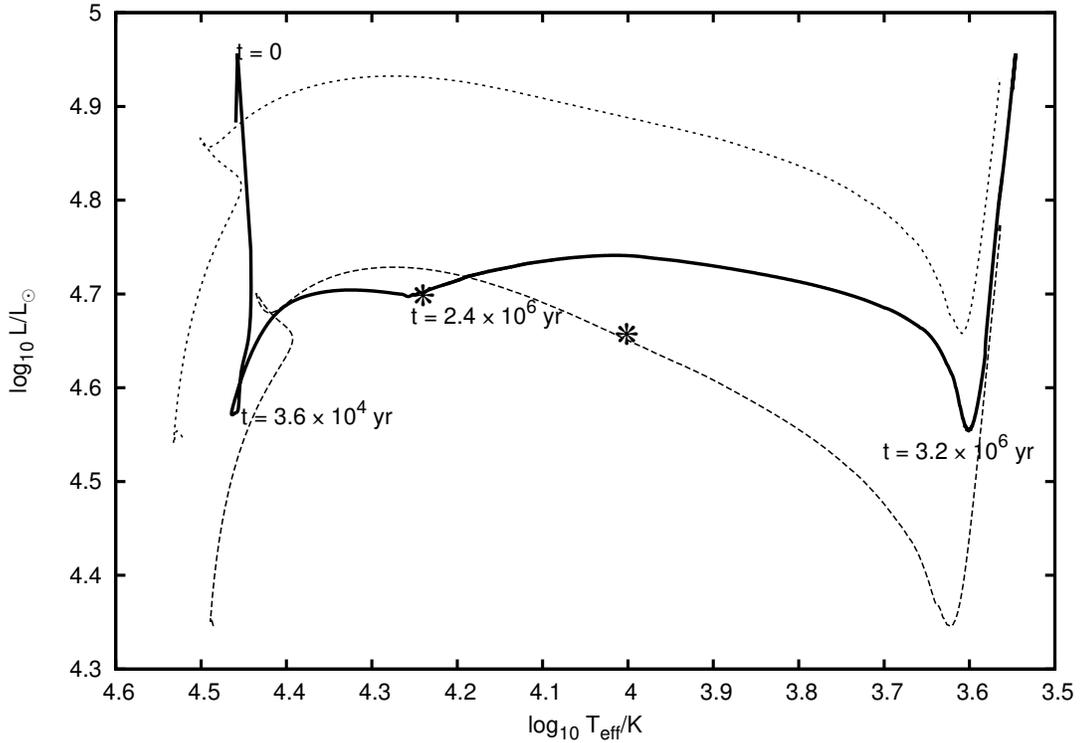}
\else
   \includegraphics[angle=270,width=\textwidth]{hrd_tams10+7-3}
\fi
\caption{Evolution track of a merger remnant from a $10$ and a $7 \mathrm{M}_\odot$
star where the primary was at the end of the main sequence. The dashed track is
a main sequence star of the same mass, the dotted track is a homogenised model.
The points of helium ignition for the collision product and the normal star are
indicated by $\ast$.
}
\label{fig:hrd_highmass}
\end{figure}
The evolution of remnants from the collision of hig mass stars (here
loosely taken to mean collisions involving stars that have $M \gtrsim 5 \mathrm{M}_\odot$)
are mainly relevant for the formation of very massive stars 
($\sim 100 \mathrm{M}_\odot$) or so-called merger runaways whereby repeated collisions
are thought to lead to the formation of a star with a mass up to $\sim 1000 \mathrm{M}_\odot$
\citep{article:portegies_zwart_runaway_merger}. Some authors have tried to 
study the evolution of such massive stars with normal single star models 
\citep{article:belkus_very_massive, article:yungelson_very_massive}
but this approach neglects the effect of mass increase
during the evolution as well as helium enhancement 
\citep{article:glebbeek_gaburov_in_prep1}.
Understanding the outcome of the first merger in such a sequence is important
for understanding the subsequent mergers with the same object 
\citep{article:gaburov_glebbeek_in_prep1}.

The study of mergers of massive stars is more complicated than that of mergers 
of low mass stars because of the importance of radiation pressure for their 
stability \citep{article:gaburov_massive_mixing}.
Qualitatively, the evolution of massive collision products is similar to that
of lower mass collision products: they settle to a main-sequence like position
on a thermal timescale and the inner region is mixed by convection and
thermohaline mixing while the envelope is not mixed.

An interesting scenario is the collision between a primary star that has just
moved off the main sequence and a lower mass star that is still on the main
sequence. The hydrogen exhausted core of the primary will end up in the centre
of the collision product, but because there is no hydrogen in the core there
is no core nuclear buring until the ignition of helium. In particular, no
hydrogen will be mixed into the core by convection: the star does not
rejuvenate. This means that the collision product will contract to a position
at the beginning of the Hertzsprung-gap rather than the main sequence, but
with the lower core mass appropriate for a lower mass star.
The remnant of the collision between a $10$ and a $7 \mathrm{M}_\odot$ solar mass
star shown in figure \ref{fig:hrd_highmass} contracts on a timescale of 
$3.6\times 10^4 \mathrm{yr}$, after which the star starts to cross the
Hertzsprung-gap. Due to its lower core mass, the star begins core helium 
burning (CHeB) while it is bluer than a normal star of the same mass and 
metallicity. Like a more massive star or a star of lower metallicity the 
collision product spends about one third of its CHeB phase 
in the blue part of the colour magnitude diagram. It takes $\sim 1 \mathrm{Myr}$
to cross from the beginning of the Hertzsprung-gap to the red branch while
an ordinary star of the same mass takes only $\sim 4 \times 10^{-2} 
\mathrm{M yr}$

A major complication in the study of massive collision products is the 
uncertainty in the mass loss rates for the remnants of these stars as well as 
the interplay between mass loss and rotation. Especially after multiple 
collisions the collision products can become very bright, close to their 
Eddington luminosity
\beq
L_\mathrm{edd} = \frac{4\pi G c M}{\kappa}
\eeq
Stars that become this bright have a high mass loss rate or eject mass in
large outbursts. A better understanding of the mass loss rates for this type
of object is needed to predict the outcome of runaway mergers
\citep{article:glebbeek_gaburov_in_prep1}.

\section{Conclusions}
Stellar collisions are a means of making stars that appear in unusual 
locations in the colour magnitude diagram, such as blue stragglers. Remnants 
from high mass 
stars that merge after the end of core hydrogen burning can become much
brighter while crossing the Hertzsprung-gap and stay there much longer.

Approximating mergers with ordinary stellar models or fully mixed models can
give significantly different results compared to proper detailed models.

\section{Unanswered Questions}
The main uncertainty in the evolution of stellar collision products is how
they lose their excess angular momentum and, consequently, the importance of chemical mixing 
due to rotational instabilities.

It is known that collision products need to lose angular momentum before they
can settle into hydrostatic equilibrium. Magnetic fields might play an
important role in this angular momentum loss, but nothing is currently known
about the configuration of the magnetic field after the merger.

Very massive collision products can reach luminosities close to their Eddington
limits, particularly for runaway mergers where many stars collide. The mass 
loss rates become very uncertain at this point. This introduces a large 
uncertainlty in the evolution of very massive stars and the type and mass of 
their remnants since these are mainly are determined by the mass loss rate.

\begin{theacknowledgments}
EG acknowledges support from NWO.
\end{theacknowledgments}

\bibliographystyle{mn2e}   % if natbib is available
\bibliography{starscoll}

\begin{thebibliography}{}

\bibitem[\protect\citeauthoryear{{Belkus}, {Van Bever} \&
  {Vanbeveren}}{{Belkus} et~al.}{2007}]{article:belkus_very_massive}
{Belkus} H.,  {Van Bever} J.,    {Vanbeveren} D.,  2007, \apj, 659, 1576

\bibitem[\protect\citeauthoryear{{Davies}, {Piotto} \& {de Angeli}}{{Davies}
  et~al.}{2004}]{article:davies_bss_formatio}
{Davies} M.~B.,  {Piotto} G.,    {de Angeli} F.,  2004, \mnras, 349, 129

\bibitem[\protect\citeauthoryear{{Eggleton}}{{Eggleton}}{1971}]{article:egglet%
on_evlowmass}
{Eggleton} P.~P.,  1971, \mnras, 151, 351

\bibitem[\protect\citeauthoryear{{Gaburov}, {Glebbeek}, {Portegies Zwart} \&
  R.}{{Gaburov} et~al.}{2007}]{article:gaburov_glebbeek_in_prep1}
{Gaburov} E.,  {Glebbeek} E.,  {Portegies Zwart} S.,    R. P.~O., , 2007, in
  preparation

\bibitem[\protect\citeauthoryear{{Gaburov}, {Lombardi} \& {Portegies
  Zwart}}{{Gaburov} et~al.}{2007}]{article:gaburov_massive_mixing}
{Gaburov} E.,  {Lombardi} J.~C.,    {Portegies Zwart} S.,  2007, ArXiv
  e-prints, 707

\bibitem[\protect\citeauthoryear{{Glebbeek}, {Gaburov} \& {Pols}
  O.~R.and~{Portegies Zwart}}{{Glebbeek}
  et~al.}{2007}]{article:glebbeek_gaburov_in_prep1}
{Glebbeek} E.,  {Gaburov} E.,    {Pols} O.~R.and~{Portegies Zwart} S., , 2007,
  in preparation

\bibitem[\protect\citeauthoryear{{Glebbeek} \& {Pols}}{{Glebbeek} \&
  {Pols}}{2007}]{article:glebbeek_pols_in_prep2}
{Glebbeek} E.,  {Pols} O.~R., , 2007, in preparation

\bibitem[\protect\citeauthoryear{{Glebbeek}, {Pols} \& {Hurley}}{{Glebbeek}
  et~al.}{2007}]{article:glebbeek_pols_in_prep1}
{Glebbeek} E.,  {Pols} O.~R.,    {Hurley} J.~R., , 2007, in preparation

\bibitem[\protect\citeauthoryear{{Heger}, {Langer} \& {Woosley}}{{Heger}
  et~al.}{2000}]{article:heger_rotation_1}
{Heger} A.,  {Langer} N.,    {Woosley} S.~E.,  2000, \apj, 528, 368

\bibitem[\protect\citeauthoryear{{Hurley}, {Pols}, {Aarseth} \&
  {Tout}}{{Hurley} et~al.}{2005}]{article:hurley_m67}
{Hurley} J.~R.,  {Pols} O.~R.,  {Aarseth} S.~J.,    {Tout} C.~A.,  2005,
  \mnras, 363, 293

\bibitem[\protect\citeauthoryear{{Hut}, {Shara}, {Aarseth}, {Klessen},
  {Lombardi} Jr., {Makino}, {McMillan}, {Pols}, {Teuben} \& {Webbink}}{{Hut}
  et~al.}{2003}]{article:modest1}
{Hut} P.,  {Shara} M.~M.,  {Aarseth} S.~J.,  {Klessen} R.~S.,  {Lombardi} Jr.
  J.~C.,  {Makino} J.,  {McMillan} S.,  {Pols} O.~R.,  {Teuben} P.~J.,
  {Webbink} R.~F.,  2003, New Astronomy, 8, 337

\bibitem[\protect\citeauthoryear{{Kippenhahn}, {Ruschenplatt} \&
  {Thomas}}{{Kippenhahn} et~al.}{1980}]{article:kippenhahn_thermohalinemixing}
{Kippenhahn} R.,  {Ruschenplatt} G.,    {Thomas} H.-C.,  1980, \aap, 91, 175

\bibitem[\protect\citeauthoryear{{Kippenhahn} \& {Weigert}}{{Kippenhahn} \&
  {Weigert}}{1990}]{book:kippenhahn_weigert}
{Kippenhahn} R.,  {Weigert} A.,  1990, {Stellar Structure and Evolution}.
Stellar Structure and Evolution, XVI, 468 pp.~192 figs..~ Springer-Verlag
  Berlin Heidelberg New York.~Also Astronomy and Astrophysics Library

\bibitem[\protect\citeauthoryear{{Livio}}{{Livio}}{1993}]{article:livio_blue_s%
tragglers}
{Livio} M.,  1993, in {Saffer} R.~A.,  ed., Blue Stragglers Vol.~53 of
  Astronomical Society of the Pacific Conference Series, {Blue Stragglers: The
  Failure of Occam's Razor?}.
pp~3--+

\bibitem[\protect\citeauthoryear{{Lombardi} Jr., {Warren}, {Rasio}, {Sills} \&
  {Warren}}{{Lombardi} et~al.}{2002}]{article:lombardi_mmas}
{Lombardi} Jr. J.~C.,  {Warren} J.~S.,  {Rasio} F.~A.,  {Sills} A.,    {Warren}
  A.~R.,  2002, \apj, 568, 939

\bibitem[\protect\citeauthoryear{{Montgomery}, {Marschall} \&
  {Janes}}{{Montgomery} et~al.}{1993}]{article:montgomery_m67}
{Montgomery} K.~A.,  {Marschall} L.~A.,    {Janes} K.~A.,  1993, \aj, 106, 181

\bibitem[\protect\citeauthoryear{{Pinsonneault}, {Kawaler}, {Sofia} \&
  {Demarque}}{{Pinsonneault} et~al.}{1989}]{article:pinsonneault_rotation}
{Pinsonneault} M.~H.,  {Kawaler} S.~D.,  {Sofia} S.,    {Demarque} P.,  1989,
  \apj, 338, 424

\bibitem[\protect\citeauthoryear{{Piotto}, {De Angeli}, {King}, {Djorgovski},
  {Bono}, {Cassisi}, {Meylan}, {Recio-Blanco}, {Rich} \& {Davies}}{{Piotto}
  et~al.}{2004}]{article:piotto_freq_bss}
{Piotto} G.,  {De Angeli} F.,  {King} I.~R.,  {Djorgovski} S.~G.,  {Bono} G.,
  {Cassisi} S.,  {Meylan} G.,  {Recio-Blanco} A.,  {Rich} R.~M.,    {Davies}
  M.~B.,  2004, \apjl, 604, L109

\bibitem[\protect\citeauthoryear{{Pols}, {Schroder}, {Hurley}, {Tout} \&
  {Eggleton}}{{Pols} et~al.}{1998}]{article:pols_evmodels}
{Pols} O.~R.,  {Schroder} K.-P.,  {Hurley} J.~R.,  {Tout} C.~A.,    {Eggleton}
  P.~P.,  1998, \mnras, 298, 525

\bibitem[\protect\citeauthoryear{{Pols}, {Tout}, {Eggleton} \& {Han}}{{Pols}
  et~al.}{1995}]{article:pols_approxphys}
{Pols} O.~R.,  {Tout} C.~A.,  {Eggleton} P.~P.,    {Han} Z.,  1995, \mnras,
  274, 964

\bibitem[\protect\citeauthoryear{{Portegies Zwart}, {Baumgardt}, {Hut},
  {Makino} \& {McMillan}}{{Portegies Zwart}
  et~al.}{2004}]{article:portegies_zwart_runaway_merger}
{Portegies Zwart} S.~F.,  {Baumgardt} H.,  {Hut} P.,  {Makino} J.,
  {McMillan} S.~L.~W.,  2004, \nat, 428, 724

\bibitem[\protect\citeauthoryear{{Portegies Zwart} \& {McMillan}}{{Portegies
  Zwart} \&
  {McMillan}}{2002}]{article:portegies_zwart_intermediate_mass_blackhole}
{Portegies Zwart} S.~F.,  {McMillan} S.~L.~W.,  2002, \apj, 576, 899

\bibitem[\protect\citeauthoryear{{Sills}, {Adams} \& {Davies}}{{Sills}
  et~al.}{2005}]{article:sills_angular_momentum}
{Sills} A.,  {Adams} T.,    {Davies} M.~B.,  2005, \mnras, 358, 716

\bibitem[\protect\citeauthoryear{{Sills}, {Lombardi} Jr., {Bailyn}, {Demarque},
  {Rasio} \& {Shapiro}}{{Sills} et~al.}{1997}]{article:sills_on_axis}
{Sills} A.,  {Lombardi} Jr. J.~C.,  {Bailyn} C.~D.,  {Demarque} P.,  {Rasio}
  F.~A.,    {Shapiro} S.~L.,  1997, \apj, 487, 290

\bibitem[\protect\citeauthoryear{{Stancliffe}, {Glebbeek}, {Izzard} \&
  {Pols}}{{Stancliffe} et~al.}{2007}]{article:stancliffe_thmixing}
{Stancliffe} R.~J.,  {Glebbeek} E.,  {Izzard} R.~G.,    {Pols} O.~R.,  2007,
  \aap, 464, L57

\bibitem[\protect\citeauthoryear{{Tout}, {Aarseth}, {Pols} \&
  {Eggleton}}{{Tout} et~al.}{1997}]{article:tout_evolution_models}
{Tout} C.~A.,  {Aarseth} S.~J.,  {Pols} O.~R.,    {Eggleton} P.~P.,  1997,
  \mnras, 291, 732

\bibitem[\protect\citeauthoryear{{Yungelson}}{{Yungelson}}{2006}]{article:yung%
elson_very_massive}
{Yungelson} L.~R.,  2006, Calibrating the Top of the Stellar M-L Relation, 26th
  meeting of the IAU, Joint Discussion 5, 16 August 2006, Prague, Czech
  Republic, JD05, \#9, 5

\end{thebibliography}

\end{document}